\documentclass[pra,twocolumn,amsmath]{revtex4}
\usepackage{graphicx}

\begin{document}

\def\ket#1{|#1\rangle} 
\def\bra#1{\langle#1|}
\def\av#1{\langle#1\rangle}
\def\myarrow{\mathop{\longrightarrow}}

\title{Observation of lasing without inversion in a hot rubidium vapor under electromagnetically-induced transparency conditions}

\author{Haibin Wu}
\author{Min Xiao}
\email{mxiao@uark.edu} 
\author{J. Gea-Banacloche}
\email{jgeabana@uark.edu}
\affiliation{Department of Physics, University of Arkansas, Fayetteville, AR 72701}

\begin{abstract}
We have observed CW lasing without inversion in a gas of hot rubidium atoms in an optical cavity under conditions of electromagnetically-induced transparency (EIT).  The medium is pumped coherently and resonantly by a single ``coupling'' beam which also produces  EIT in the lasing transition.  The steady-state intensity exhibits thresholds as a function of the atomic density and the strength of the coupling beam.  A theoretical model for an effective three-level lambda system indicates that gain without inversion is possible in this system if the two ground states are coupled by depolarizing collisions, and if the decay branching ratios meet certain conditions.  
\end{abstract}

\maketitle

Although the theoretical possibility of lasing without inversion (LWI) has been known for almost two decades \cite{vskaya,harris,scully1}, the number of experimental demonstrations of continuous-wave laser action without population inversion is still relatively small \cite{zibrov,padmabandu}.  Here, we report the observation of lasing without inversion in a hot gas of rubidium atoms under conditions of electromagnetically-induced transparency (EIT) \cite{harris2}, in what we believe is the simplest setup considered to date, with only one external laser beam providing the pumping and, simultaneously, the EIT ``coupling'' field.  We also present a simple theoretical model, involving only three atomic levels in the lambda configuration, where the two ground states are connected by incoherent decay rates, such as might arise from depolarizing collisions in the very dense, hot gas.

The experimental setup, schematically represented in Figure 1, is very similar to the one in which we recently observed EIT and the transmission spectrum associated with bright polaritons in the strong-coupling regime \cite{us}, only now no probe beam is injected into the optical ring cavity.  This consists of three mirrors: the (in the former experiment) input mirror M1 and output mirror M2 have $3\%$ and $1.4\%$ transmissivities, respectively, and M3 is a high reflector mounted on a PZT for cavity frequency scanning and locking (giving an empty cavity linewidth of about 17 MHz, which corresponds to a finesse of about 48). The cavity length is 37 cm.  The rubidium vapor cell is 7 cm long, with Brewster windows; it is wrapped in  $\mu$-metal sheets to shield from external magnetic fields and a heat tape is placed outside the  $\mu$-metal sheets for controlling the temperature (which determines the atomic density).  The pump laser, a commercial high-power CW diode laser, is frequency stabilized and locked to the atomic transition $5S_{1/2}, F=2 \to 5 P_{1/2}, F^\prime = 2$  of the D1 line of $^{87}$Rb, and has a linewidth of about $1.0$ MHz. It is carefully aligned through the vapor cell at a small angle ($\approx 2^\circ$) from the cavity axis, and with s-polarization, thereby avoiding to be resonant with the optical cavity. 
\begin{figure}
\includegraphics[height=5cm]{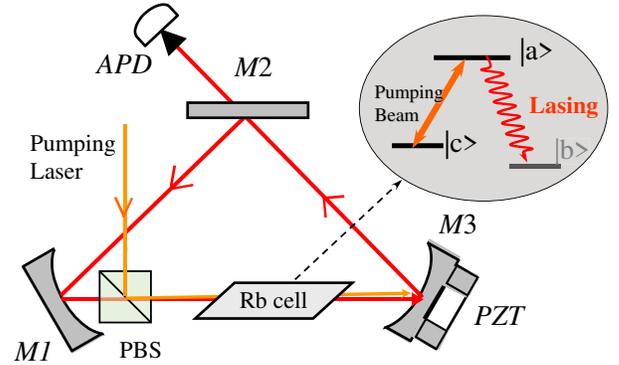}
\caption[example] 
   { \label{fig:fig1} 
The experimental setup.  Inset: simplified three-level model (ignoring the Zeeman magnetic sublevels).}
\end{figure} 
The optical depth (OD) of the atomic medium can be calculated using
\begin{equation}
OD = N\sigma l = N\, \frac{3 \lambda^2}{8 \pi n^2}\,\frac{\gamma}{\Delta\Omega_D}\, l,
\label{e1}
\end{equation}
where $N$ is the atomic density; $n$ is the refractive index; $\gamma$  is the decay rate of the upper level of the atom, and $\Delta\Omega_D$ is the Doppler width, respectively. At $90^\circ$C the Doppler width is 552 MHz, which gives an OD of 139.  The atomic medium should be completely opaque for the transition corresponding to $\ket a\to \ket b$ at such high OD, but due to the existence of EIT, the p-polarized generated signal beam at this transition has low absorption and can pass through the atomic medium and oscillate in the cavity when the cavity is scanned (due to the closeness in frequency between pump and signal beam, the first-order Doppler broadening is eliminated when they propagate in the same direction through the cell \cite{jgb}).

As the pumping field power increases, a small output signal, at a frequency corresponding to the $F^\prime = 2 \to F = 1$ transition, starts to emerge from the cavity, and builds up quickly as the pump power further increases. A typical cavity output result is shown in Fig. 2(a) at a modest pump power ($21.8$ mW). The linewidth of the output peak is $3.30\pm 0.22$ MHz, which is far narrower than the atomic linewidth and cavity linewidth. When the intensity of the output signal is measured as a function of pumping power, a clear threshold behavior can be seen, as shown in Fig. 3. As the pumping power grows, the output signal eventually peaks and starts to decrease, an effect that can be predicted from the theoretical model developed below. The mark (cross) on Fig. 3 indicates the pump power used in Fig. 2(a). 
\begin{figure}
\includegraphics[height=6cm]{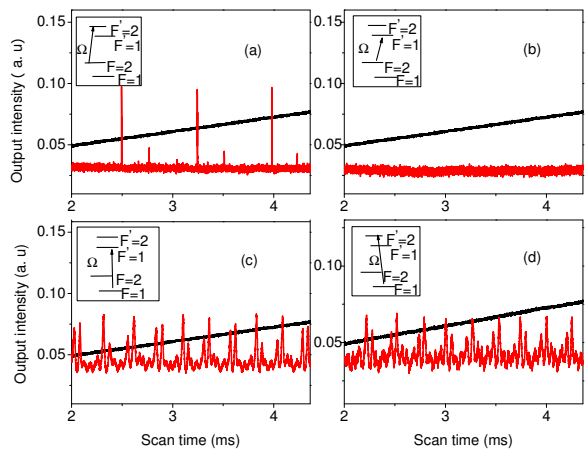}
\caption[example] 
   { \label{fig:fig2} 
Results of cavity scans when the coupling (pump) beam is tuned to various different transitions of $^{87}$Rb.  The temperature is $T=90^\circ$C, and the pump power is $21.8$ mW, corresponding to a coupling Rabi frequency of 148 MHz.}
\end{figure} 
The second peak visible in Fig.~2(a) is at the frequency of the transition $F^\prime = 2 \to F = 2$, which is the same as the pump field, only it has a different (p) polarization and propagates at a small angle to it, so as to circulate in the cavity.  It is most likely due to LWI (and EIT) among Zeeman sublevels of the $F^\prime = 2$ and $F = 2$ states.  We also have observed that as the pumping power gets large, more peaks show up at different frequencies, which may be the result of four-wave mixing and other Raman processes in the medium. We hope to investigate this in detail in a future study. 
\begin{figure}
\includegraphics[height=6cm]{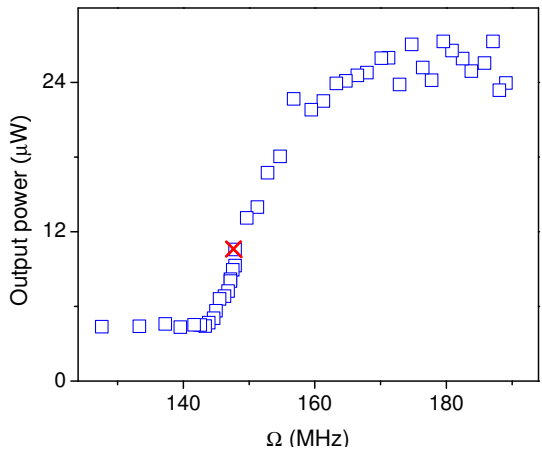}
\caption[example] 
   { \label{fig:fig3} 
The threshold behavior of the emitted light as a function of the coupling beam intensity.  The temperature is $T=90^\circ$C, corresponding to an optical depth of 139.}
\end{figure} 
It is interesting to notice that when the pump field is tuned to the $5S_{1/2}, F=2\to 5 P_{1/2}, F^\prime = 1$ transition, no lasing action has been observed even with careful cavity and pump field alignments, as shown in Fig. 2(b). For two other transitions (i.e., $5S_{1/2}, F=1\to 5 P_{1/2}, F^\prime = 1$ and $5S_{1/2}, F=1\to 5 P_{1/2}, F^\prime = 2$) with the pump beam at same power, only fluorescent background (incoherent scattering) was observed, which could not lead to lasing even when the pumping power was further increased to a high value (all the way to the limit of our diode laser at 50 mW). 

Another interesting feature of this system is the dependence of the lasing characteristics on the atomic density or OD. At low temperature (small OD), lasing does not appear even at very high pumping power.  For a given pumping power of 25 mW, the measured cavity output power as a function of OD (up to OD=326) is presented in Fig. 4, where a clear threshold behavior may be seen. This is to be expected, since in order to have laser oscillation the atomic gain, which is proportional to the atomic density, has to exceed the cavity losses.  This threshold value depends critically on the pumping power, and can be lowered for higher pumping power.  The eventual saturation of the output power with increasing atomic density can in fact be predicted (qualitatively) from the simple three-level model discussed below.
\begin{figure}
\includegraphics[height=6cm]{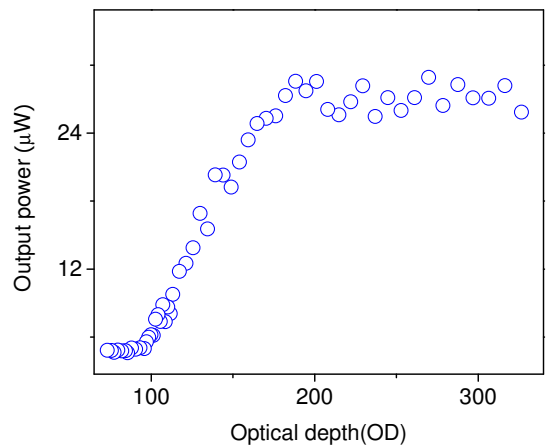}
\caption[example] 
   { \label{fig:fig4} 
The threshold behavior of the emitted light as a function of the atomic density. The pumping power is fixed at 25 mW ($\Omega = 156$ MHz). The maximum OD shown (326) corresponds to a temperature of $103^\circ$C.}
\end{figure} 

Our system differs from the previous experiments that have shown steady-state LWI oscillation \cite{zibrov, padmabandu} in that it does not require an incoherent pump in addition to the EIT coupling beam.  Conceptually, it is rather related to Raman lasers (see, e.g., \cite{eschmann}, and references therein).  Unlike in parametric-gain schemes, our pump beam is resonant with the atomic transition, so steady-state population is actually transfered to the upper level $\ket a$.  Another difference with most Raman-gain schemes is that, at the temperatures at which our system operates, the populations of both ground states (ignoring degeneracies) should be essentially the same.  When degeneracy is considered, one expects the $\ket c$ ($F=2$) manifold to have initially $5/3$ the total population of the $\ket b$ ($F=1$) manifold; hence, even if the pump field was strong enough to bleach the $\ket c\to \ket a$ transition, that would still make the population of state $\ket a$ equal to only $5/6$ that of the lower lasing state $\ket b$, so there would not be population inversion in the $\ket a \to \ket b$ transition in any case.  Finally, note also that in order to prevent the lasing field from optically pumping the whole system to the state $\ket b$ (and thus shutting itself off), we need to postulate some mechanism to feed population back from $\ket b$ to $\ket a$.  We believe that this mechanism could be provided by depolarizing collisions between atoms \cite{franzen}.

With this insight, we can develop a ``toy model'' that predicts the possibility of steady-state gain without inversion in the simplified three-level system shown in the inset in Figure 1.  Take the coupling beam to be on resonance with the $\ket c \to \ket a$ transition and to have a Rabi frequency $\Omega$.  If the cavity field has an amplitude $a$ and is resonant with the $\ket b \to \ket a$ transition, the single-atom equations of motion are
\begin{align}
\frac{d}{dt}i\rho_{ab} &= - g a \left(\rho_{aa}-\rho_{bb}\right) + \frac{\Omega}{2} \rho_{cb} - \gamma_{ba} i \rho_{ab} \cr
\frac{d}{dt} \rho_{cb} &= g a i \rho_{ca} - \frac{\Omega}{2} i \rho_{ab} -\gamma_{bc}\rho_{cb} \cr
\frac{d}{dt}i\rho_{ca} &= -\frac{\Omega}{2}\left(1-\rho_{bb}-2\rho_{aa}\right) - g a \rho_{cb} - \gamma_{ac} i\rho_{ca} \cr
\frac{d}{dt} \rho_{aa} &= -\Omega i \rho_{ca} + 2 g a i\rho_{ab} - \gamma_a \rho_{aa}\cr
\frac{d}{dt}\rho_{bb} &= -2 g a i\rho_{ab} + f \gamma_a \rho_{aa} - \gamma_b \rho_{bb} + \gamma_c \left(1 - \rho_{aa}-\rho_{bb}\right)\cr
\label{e2}
\end{align}
Some simplifications have already been used in (\ref{e2}), such as the closed system assumption, $\rho_{cc} = 1 - \rho_{aa}-\rho_{bb}$ and the fact that, on resonance $i\rho_{ab}$ and $i\rho_{ca}$ are real, so $i\rho_{ba} = -i\rho_{ab}$, etc.  We assume that collisions cause a flow of population from $\ket c$ to $\ket b$ at a rate $\gamma_c$, and from $\ket b$ to $\ket c$ at a rate $\gamma_b$.  Then the decay rate of the coherence $\gamma_{bc}$ must satisfy
\begin{equation}
\gamma_{bc} \ge \frac{\gamma_b + \gamma_c}{2} 
\label{e3}
\end{equation}
We are also assuming that a fraction $f$ of the spontaneous decay of level $\ket a$ (with overall decay rate $\gamma_a$) ends up in $\ket b$; by the closed-system assumption, this means that a fraction $1-f$ ends up in $\ket c$.  

The steady-state linear gain is given by the coefficient of $a$ in a power series expansion of $-ig N(\rho_{ab})_{ss}$, where $N$ is the atomic density, which from Eqs.~(\ref{e3}) is easily calculated to be
\begin{widetext}
\begin{equation}
G_{ss}^l = 2g^2 N\frac{\Omega^2\left(\gamma_a(\gamma_b- 2 f \gamma_{bc})+2\gamma_{bc}(\gamma_b-\gamma_c)\right) -4\gamma_a\gamma_{ac}\gamma_{bc}\gamma_c }{ (\Omega^2 + 4 \gamma_{ba}\gamma_{bc})(\Omega^2(f\gamma_a+2\gamma_b+\gamma_c)+2\gamma_a\gamma_{ac}(\gamma_b+\gamma_c))} 
\label{e4}
\end{equation}
\end{widetext}
This will be positive for large enough $\Omega$, {\it if\/} the coefficient of $\Omega^2$ is positive.  Since we expect $\gamma_a$ to be much larger than $\gamma_b,\gamma_c$, or $\gamma_{bc}$, the important term is the first one, that is, we require
\begin{equation}
\gamma_b > 2 f \gamma_{bc}
\label{e5}
\end{equation}
If Eq.~(\ref{e5}) and (\ref{e3}) are taken together, it is clear that we require 
\begin{equation}
f< \frac{\gamma_b}{\gamma_b+\gamma_c} < 1.
\label{e6}
\end{equation}
Interestingly, this condition cannot be satisfied for both ``legs'' of the $\Lambda$ system.  If, as in Fig. 2(c), we switch the roles of states $\ket b$ and $\ket c$, then the condition becomes $1-f <  \gamma_c/(\gamma_b+\gamma_c)$, which is equivalent to $f > \gamma_b/(\gamma_b+\gamma_c)$, i.e., the opposite of (\ref{e6}).  This explains why we do not see lasing in the configuration of Fig.~1(d).  Note, however, that (\ref{e6}) is only a necessary, but not a sufficient, condition; the necessary and sufficient condition is (\ref{e5}), and it could very well be the case that both $\gamma_b > 2 f \gamma_{bc}$ and $\gamma_c > 2 (1-f) \gamma_{bc}$ are violated, so there is no gain on \emph{either} configuration (as Figs.~1(b) and 1(c) show).

One can also solve (\ref{e2}) for the steady-state population inversion.  Neglecting saturation, that is, to lowest order in the cavity field $a$, the result is
\begin{equation}
\rho_{aa} - \rho_{bb} = - \frac{2\gamma_a \gamma_c\gamma_{ac} + \Omega^2 (f \gamma_a + \gamma_c - \gamma_b)}{2 \gamma_a \gamma_{ac}(\gamma_b+\gamma_c)+ \Omega^2(f \gamma_a+2 \gamma_b+\gamma_c)}
\label{e7}
\end{equation}
Since it is reasonable to expect that $f\gamma_a  \gg \gamma_b$, this will always be negative, confirming the expectation that (\ref{e4}) is, indeed, inversionless gain.

A detailed calculation of the actual gain for our experimental system would have to include many effects that have been ignored in the above model, and in particular the whole Zeeman sublevel structure of the system (see, for an example, \cite{ling}), since branching ratios appear to play such a critical role.  Also, for small $\Omega$, the Doppler broadening may also have to be considered, although for $\Omega$ sufficiently large for EIT to be appreciable we may expect the gain near line center to be given by the homogeneously-broadened formulas \cite{jgb}.  Lastly, transient effects may also be important, since we estimate (based on \cite{franzen}) that the depolarizing decay rates $\gamma_b$ and $\gamma_c$ in our system may be of the order of $0.01$ MHz, so it may take an atom a time of the order of tens of microseconds to fully reach steady state, and this is comparable to the time it may take the atoms to diffuse out of the pump beam region (beam waist $\sim 100 \mu$m).  We note that transient effects in LWI have been recently investigated in \cite{scully2}. 

In any case, a very rough estimate of the gain can be obtained from (\ref{e4}) by assuming $\Omega$ is greater than all the decay rates and neglecting all the small terms, which yields $G_{ss}^l \sim 2 g^2 N \gamma_b/\Omega^2$.  In Ref. \cite{franzen} a value for the effective cross-section for exchange collisions in $^{87}$Rb of 7 to $10\times 10^{-14}$ cm$^2$ is quoted (based on an unpublished report by Dicke, Carver, Alley and Van der Ven), from which a collision depolarization rate equal to $N\sigma_{ex}\bar v_{rel}$ could be inferred, where $\bar v_{rel}$ is the relative velocity of the colliding atoms.  At $90^\circ$ we obtain, from the data in \cite{steck}, a density $N= 2.4\times 10^{18}$ m$^{-3}$, and a thermal velocity of 263 m/s, from which we estimate $\gamma_b \sim 0.013$ MHz.  Using then $\Omega \sim 160$ MHz and $g\sqrt N \sim 3$ GHz (calculated using the above density), we get from $2 g^2 N \gamma_b/\Omega^2 \simeq 9$ MHz the correct order of magnitude for the gain (recall that the cavity amplitude decay rate is of the order of 8 MHz).  Note that, even though this expression for the linear gain, $G_{ss}^l$, decreases for increasing $\Omega$, the actual steady-state intensity needs to be calculated from the saturated gain, and when this is done one indeed finds a result qualitatively similar to Fig. 3 (including the observed saturation behavior), for appropriate choices of the parameters.  

More specifically, for sufficiently large $\Omega$ the nonlinear gain in the model (\ref{e1}) is approximately given by 
\begin{equation}
G_{ss} \simeq 2 g^2 N \frac{\Omega^2(\gamma_b - 2f \gamma_{bc}) - 4 g^2 a^2 \gamma_c}{f\Omega^4 + 4 g^2 a^2 \Omega^2 + 16 g^4 a^4 (1-f)}
\label{e8}
\end{equation}
By setting this equal to the cavity decay rate $\gamma$ and solving for the steady state intensity (proportional to $a^2$) one obtains results qualitatively similar to figures 3 and 4, i.e., the intensity eventually peaks as a function of $\Omega$ and saturates as a function of $N$ (if one additionally assumes, as above, that the rates $\gamma_b$, $\gamma_c$ and $\gamma_{bc}$ are proportional to $N$).  The threshold behavior as a function of $\Omega$ is not very well captured by (\ref{e8}), but, as pointed out above, for small $\Omega$ it is probably not a good approximation to ignore the Doppler-broadening of the medium \cite{jgb}.

In conclusion, we have observed lasing without inversion in an EIT lambda system, by a mechanism that does not appear to have been considered before, where atomic collisions provide a coupling between the ground states.  A single laser beam provides both the pumping and the necessary medium transparency, making this probably the simplest LWI setup yet demonstrated. 

Partial support from the National Science Foundation is gratefully acknowledged.

\end{document}